# Framework for Incorporating Community Social Vulnerability in the Assessment of Hurricane-Induced Wind Risk to Residential Buildings

S. Amin Enderami [a], Ram K. Mazumder [b], Elaina J. Sutley [c]

[a] University of Kansas, Lawrence, KS, USA, a.enderami@ku.edu
[b] University of Kansas, Lawrence, KS, USA, rkmazumder@ku.edu
[c] University of Kansas, Lawrence, KS, USA, enjsutley@ku.edu

ABSTRACT: The damage to residential buildings caused by hurricane-induced hazards is significant in the United States. There is a substantial body of research related to assessing physical damages of residential buildings and their direct economic impacts, yet the social consequences of such damages have generally been neglected. Since more socially vulnerable community members tend to live in more vulnerable buildings and neighborhoods, it is imperative to incorporate social vulnerability into community resilience assessment models. In this paper, a novel framework is proposed to create hurricane-induced wind risk maps that include both physical and social vulnerability, so that resilience-informed decisions can be made.

KEYWORDS: hurricane-induced wind, social vulnerability, physical vulnerability, risk analysis.

## 1 INTRODUCTION

Hurricanes are multi-hazard events that induce strong winds, storm surges, and sometimes heavy waves in coastal communities, as well as flash flooding and riverine flooding in inland communities. These hazards, together, annually cause billions of dollars in losses and civilian casualties in the coastal communities of the United States (Mazumder et al. 2021). The increasing number of such events (frequency) and their destructive impacts (intensity), as the evident consequences of climate change (Bell et al. 2018; Trenberth 2018), indicate the significance of developing community-level risk mitigation models for hurricane-induced hazards (Daniel et al. 2022). This study focuses on the impacts of hurricane-induced winds on coastal communities' residential buildings. Several studies can be found in the literature regarding improving the wind loads models (e.g.; (Guo and van de Lindt 2021; Hu et al. 2012)) and developing methods and software for examining their impacts on the built environment (e.g.; (Bezabeh et al. 2020; Simiu and Yeo 2019)). However, the consequences of hurricanes are not limited to such physical destructions and economic losses. Hurricane-induced winds often bring about substantial disruptions in social and community services (Enderami et al. 2021). Such disruptions amplify existing social disparities must be elevated in risk reduction plans to support equity. Social impacts are a function of the community's social vulnerability (Logan et al. 2021). Resilience studies prove that socially vulnerable community members often suffer more from natural hazards (Bergstrand et al. 2015).

  Defining risk as a function of hazard exposure and vulnerability, this study combines the physical vulnerability of residential buildings and the social vulnerability of residents to assess the risk of hurricane-induced winds to residential buildings located in hurricane-prone regions. Onslow County, North Carolina is selected as the study area and a wind hazard model is created according to the county's historical hurricane records. The community's residential building inventory is created using real open-source county-level data. HAZUS wind-damage fragility models are used to assess the physical vulnerability of residential buildings. Social vulnerability is measured using a novel, scalable composite score. The score is the result of combining specified demographic indicators from U.S. Census data that proxies the community's social vulnerability. Finally, a risk assessment matrix is employed to integrate social and physical vulnerabilities. The novel aspect of the proposed framework is the ability to generate community-level wind risk maps that incorporate social and physical vulnerability. These maps would allow community planners and decision-makers to make more equity-informed and risk-based decisions for enhancing community resilience.

## 2 METHODOLOGY

A novel framework is presented in this paper to capture physical and social vulnerability in developing the community-level wind risk maps for residential buildings. The framework has three main components, namely, physical vulnerability, social vulnerability, risk analysis. The physical vulnerability component assesses the physical damages of residential buildings due to exposure to hurricane-induced strong winds by employing fragility-based vulnerability functions. The output of this component would be the probability of exceeding physical damage from a predefined damage state. These probability values, later, will be mapped to corresponding physical vulnerability likelihoods within the risk analysis component to create the physical vulnerability likelihood map. The social vulnerability component uses demographic data to assign a vulnerability level to the community members. In the risk analysis component, the results from the physical vulnerability component are mapped to the social impact descriptors to account for the spatial distribution of social consequences arising from physical damage to community members who reside in that level of social vulnerability. Due to the reliability of the census data, the social vulnerability analysis should be conducted at a resolution higher than the census block group level, while the physical vulnerability component estimates damage at the individual building level.

The risk analysis component, finally, uses Equation 1 to combine the physical damage likelihood and the social impact index and calculate the risk. The estimated risk values then are ranked and interpreted by applying a widely approved risk rating scale method. Figure 1 shows a schematic representation of the proposed framework.

$$Risk = Likelihood\ of\ a\ phyical\ damage\ \times\ Social\ impcats\ of\ the\ given\ damage \qquad (1)$$

## 3 EXAMPLE COMMUNITY: ONSLOW COUNTY, NORTH CAROLINA

Onslow County, NC is a coastal county in the U.S. with a history of experiencing major hurricanes. The county comprises the City of Jacksonville, which is the County seat, and multiple towns. As of the ACS 2014-2019, 195,069 people including 64,386 households with a median income of $50,278 resided in 52,487 residential buildings in the county. The demographics match closely with U.S. national averages and provide diverse social vulnerability levels for our research purposes. A virtual testbed based on the real data of Onslow County is constructed. The testbed's community module includes a residential building inventory for physical vulnerability evaluation and a block group-level social vulnerability indicator for social vulnerability assessment.

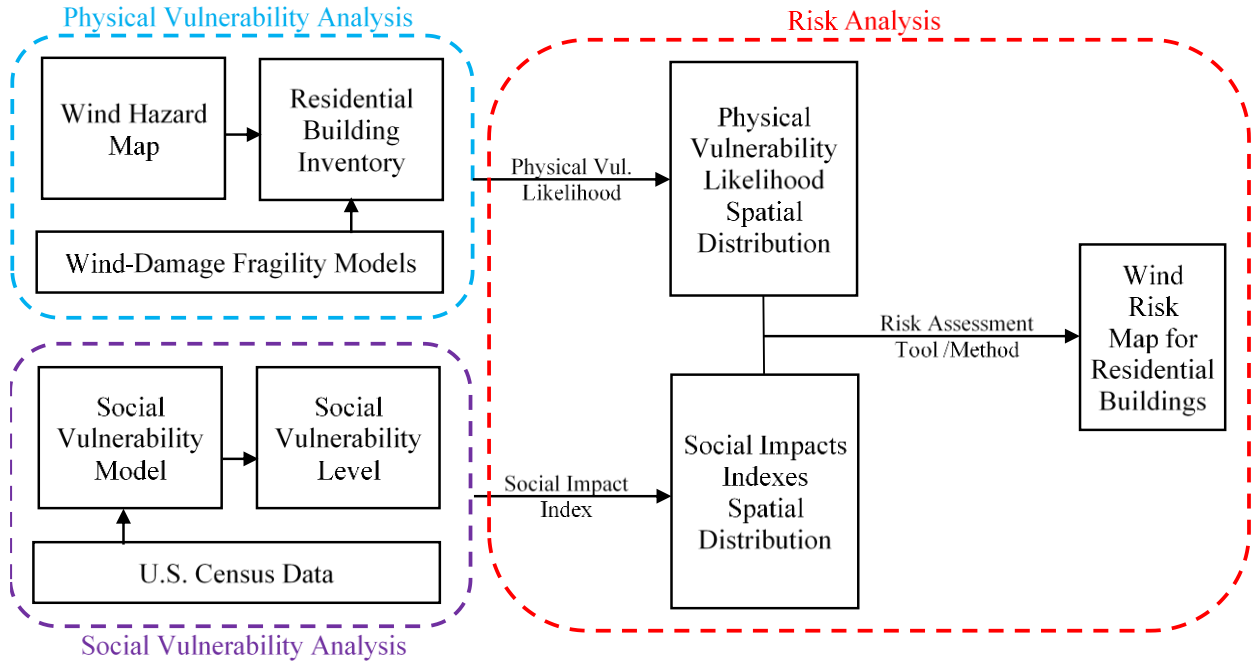

Figure 1. Schematic of the proposed framework.

The testbed has never experienced a Category 5 hurricane, however, between 1857 and 2020, three Category 4 hurricanes, including Helene (1958), Diana (1984), and Hazel (1954), were recorded within 100 km of the county, according to the NOAA historical hurricane tracks database. Figure 2 shows the trajectory of these hurricanes. The most powerful of these hurricanes, Hurricane Helene (1958), was chosen as the scenario event for the testbed's hazard module (Mazumder et al. 2021). The data required for simulating the scenario winds including location, maximum wind speed, and central pressure of the hurricane eye were obtained from the Atlantic hurricane database, known as HURDAT2. The model can estimate the wind gust speed at the location of each residential building. The assumed scenario generates the extreme value of wind gust speed during the past 165 years. Based on statistics of extreme events theory, the probability that the largest value among the $n$ previously observation will be exceeded in $N$ subsequent future observations is:

$$P(X_N > y_n) = \frac{N}{N+n} \qquad (2)$$

where $y_n$ is the maximum value among the $n$ previous observation and $X_N$ is the largest value in $N$ future subsequent observation (Ang and Tang 1984). Thus, using Equation 2, the probability that such wind gusts will be exceeded during the next 50 years is 23.5%, considering 50 years to be the typical design life for residential buildings. These values are correspondent to speeds due to a scenario hurricane with an annual exceedance probability equal to 0.0053 (MRI = 190 years). The ASCE Prestandard for Performance-Based Wind Design specifies "*operational*" and "*continuous occupancy with limited interruption*" as minimum performance objectives for Risk Category II buildings subjected to 10-year and 700-year MRI winds, respectively. According to the damage state definitions for residential buildings found in Hazard-MH (2012), these two performance objectives were mapped to "Minor" and "Severe" damage states. Thus, it is expected that damage to residential buildings, as the Risk Category II structures, will not exceed the "Moderate" level when subjected to the wind with 190 years return period.

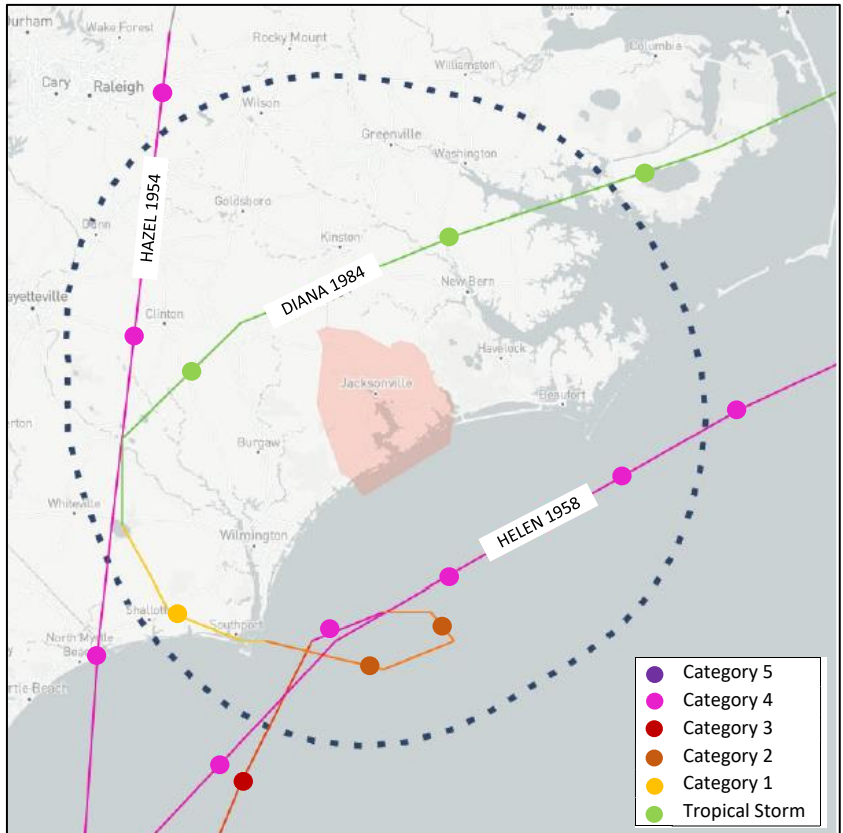

Figure 2. Historical Category 4 hurricanes recorded in the Onslow County area

## 4 RESULTS

### 4.1 *Physical Vulnerability Assessment*

The testbed's building inventory includes geospatial data on key physical attributes of the existing residential buildings portfolio as well as their correspondent fragility-based vulnerability functions. The required information to build the residential building inventory is mostly obtained from open data provided by the local government and other publicly available online resources such as Open Street Map, and tax records. The physical attributes incorporated in the inventory consist of buildings' location, dwelling type, the number of stories, exterior wall material, year built, and square footage. The accuracy of some of these data was verified using other public datasets such as Microsoft Building Footprint data (Microsoft 2020) and ReferenceUSA database (Infogroup n.d). The concept of building portfolio (Nofal and van de Lindt 2020) is applied to create the community's residential building inventory to fill the gaps in the required data. A set of 19 archetypes were mapped to the buildings in the inventory based on their dwelling type, the number of stories, construction material, and roof shape. A value of 0.35 m is taken for terrain surface roughness according to the testbed's topography. These data are used to determine the required parameters for assigning the proper HAZUS wind-damage fragility models (HAZUS-MH 2012) to each building to assess physical vulnerability. The probabilistic performance of the buildings is estimated from the probability of exceeding moderate damage state given the peak gust wind speed at each building's location. The calculated values are used to categorize the likelihood of damage into one of five categories: rare, unlikely, possible, likely, and certain. Figure 3 shows the spatial distribution of residential buildings' physical damage likelihood at the building level alongside the thresholds used to map the probability of damage exceedance to physical vulnerability likelihood categories.

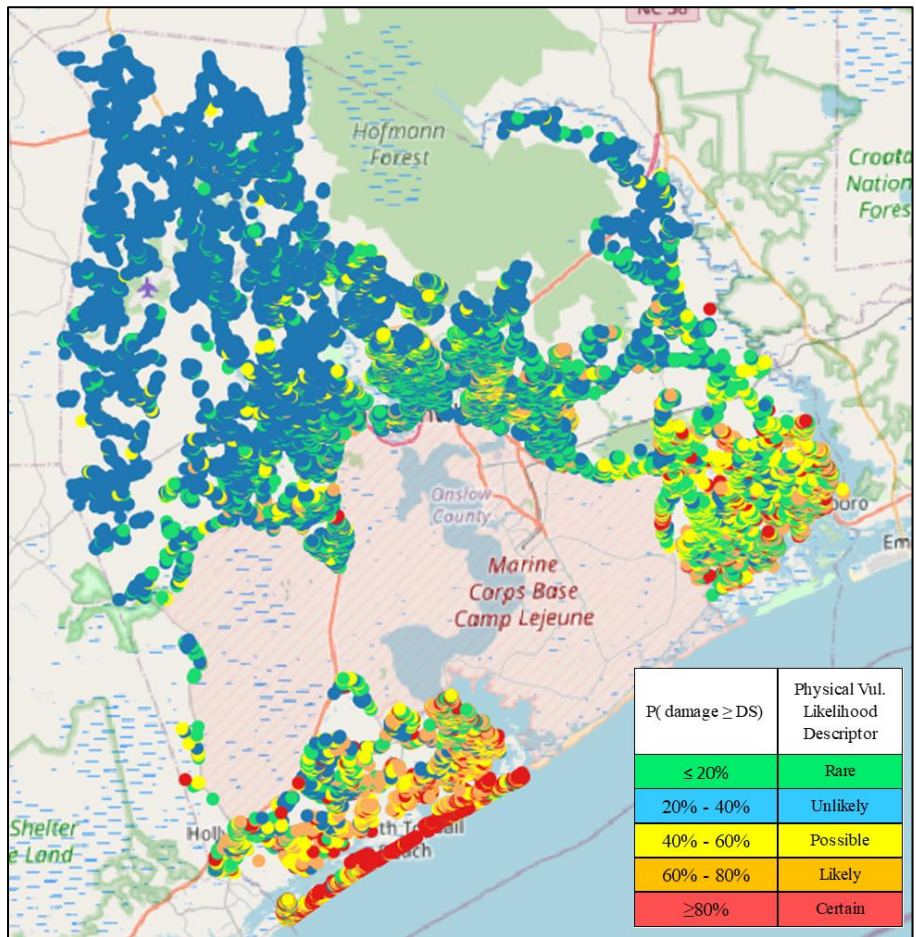

Figure 3. Onslow County residential building physical vulnerability map.

## 4.2 *Social Vulnerability Assessment*

To assess a multi-aspect qualitative feature such as community social vulnerability, scholars have been using various types of indicators. The scale and complexity of the selected indicator depend on the analysis resolution as well as the availability of data. In this study, a novel Social Vulnerability Score (SVS) developed by the authors is employed and applied to assess social vulnerability. The SVS uses U.S. Census data to estimate the demographic variables, listed in Table 1, at the location of interest at either a Census Tract or Block Group spatial scale. The variables are measured in terms of capacity ratios, i.e. the percentage of the non-vulnerable population, where zero represents complete vulnerability while 1.0 represents absolute invulnerability.

Table 1. Demographic variables used by the SVS

| No. | Description | Notation |
| --- | --- | --- |
| 1 | Percentage of white alone (not Hispanic or Latino) population | DF1 |
| 2 | Percentage of owner-occupied housing units | DF2 |
| 3 | Percentage of population earning greater than official poverty threshold | DF3 |
| 4 | Percentage of persons over age 25 with high school diploma or higher education | DF4 |
| 5 | Percentage of the population between 18 and 65 years old without disability | DF5 |

The SVS, then, calculates the ratio of the estimated variables against their national average values and aggregates them using Equation 3 while each ratio is equally weighted:

$$SVS = \frac{1}{5}\sum_{i=1}^{5} R_i \qquad (3)$$

where $R_i$ represents the calculated ratios corresponding to the demographic variables in Table 1.

The SVS maps to five levels, called zones, ranging from very low vulnerability (zone 1) to very high vulnerability (zone 5) using a standard deviation classification approach. The standard deviation is estimated considering the fact that 99.7% of values following a normal distribution lies within 3 standard deviations of the mean. Figure 4 shows the social vulnerability map at the Census Block Group (BG) level for Onslow County. To describe the social consequences of the physical damage, five social impact indexes are defined, namely low, minor, moderate, major, and catastrophic. Each Census BG is assigned an index using the mapping key shown in Figure 4. The social impact descriptor illuminates while given physical damages cause Low social impacts on a zone 1 residential building, the similar damages to a zone 5 building can have Catastrophic consequences.

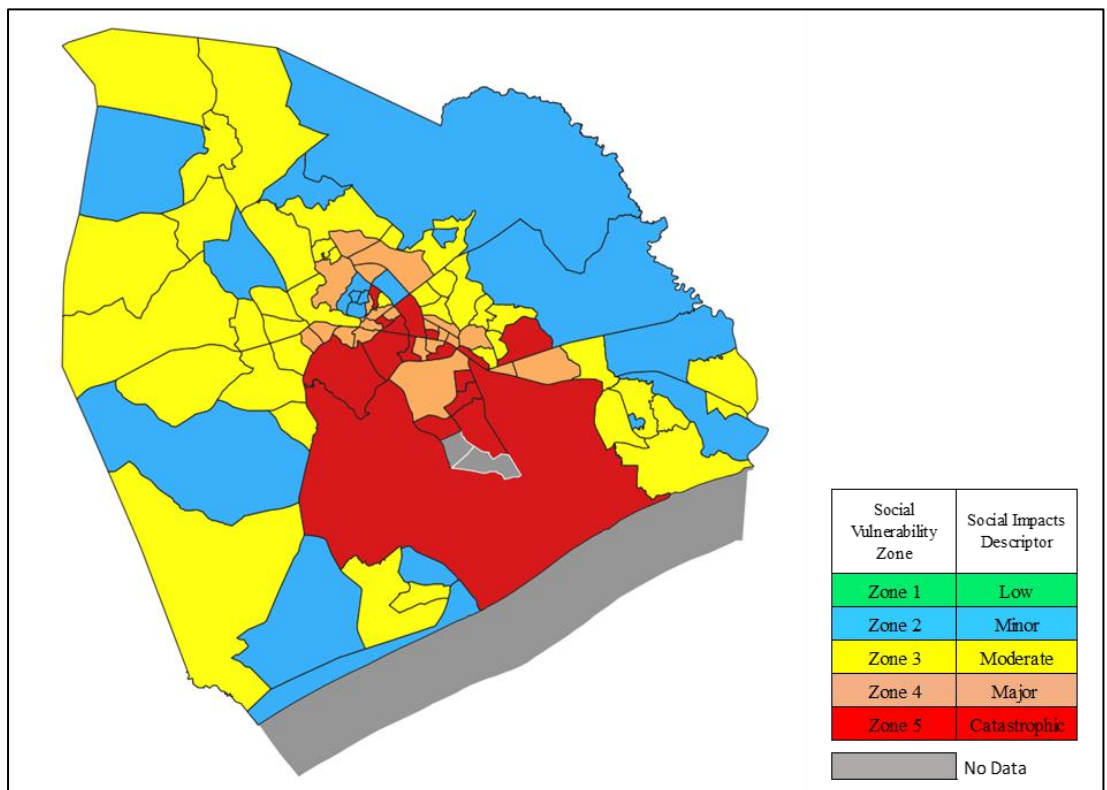

Figure 4. Onslow County BG-level SVS map.

### 4.3 *Risk Analysis*

The overall risk is calculated at the Census BG level using Equation 1 and ranked based on the risk assessment matrix shown in Figure 5. To calculate the physical damage index for a Census BG, the average of physical vulnerability likelihoods for residential buildings resided in that BG is taken. In Figure 6, Onslow County's overall wind risk map for residential buildings is shown.

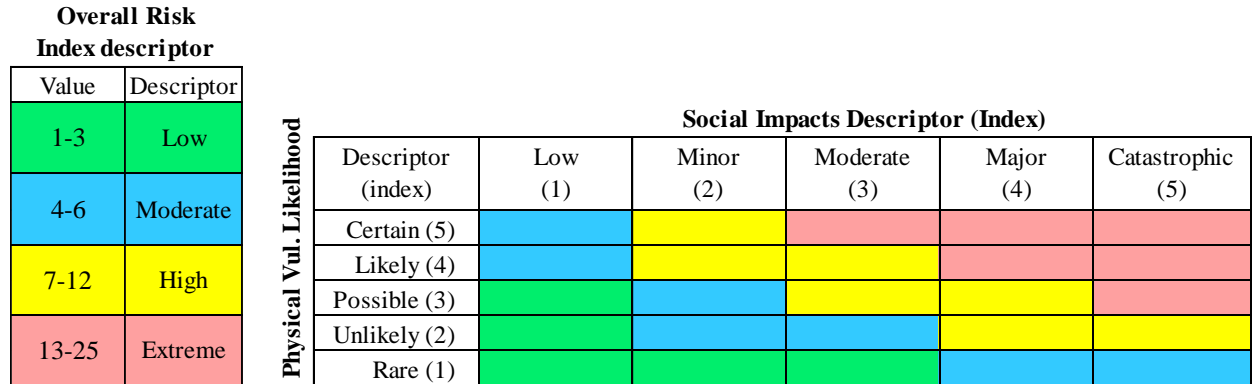

Figure 5. Risk assessment matrix

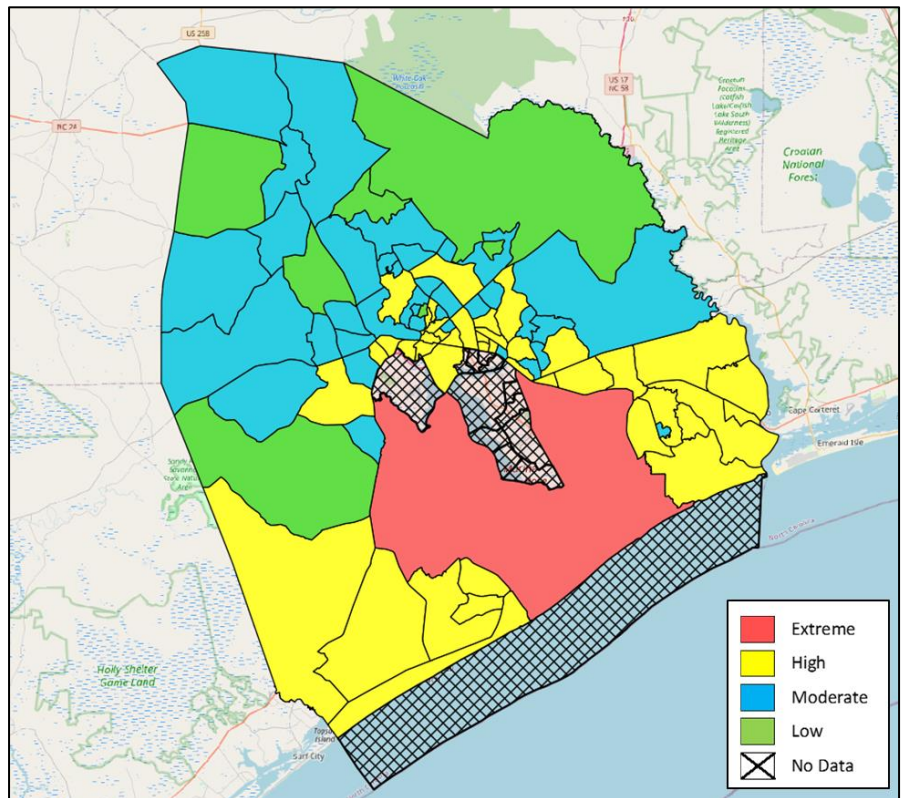

Figure 6. Onslow County overall risk map

We also compared the overall risk map with the spatial distributions of physical damage across a selection of four BGs in the example community, shown in Figure 7. The multi-color dots in Figure 7(a) represent the likelihood of moderate physical damage. The majority of the residential buildings in all four BGs are unlikely or rare to experience the predefined damage state since most of the dots are tagged blue and green. Thus, if only physical damage is included in the risk assessment, all four BGs would be at a low-risk level. However, as shown in Figure 7(b), the overall risk is different in the selected BGs because of the difference between the social vulnerability of their population (see Figure 4). The findings of this study demonstrate how equity can be prioritized by including the social vulnerability impacts in community-level natural hazards risk assessment.

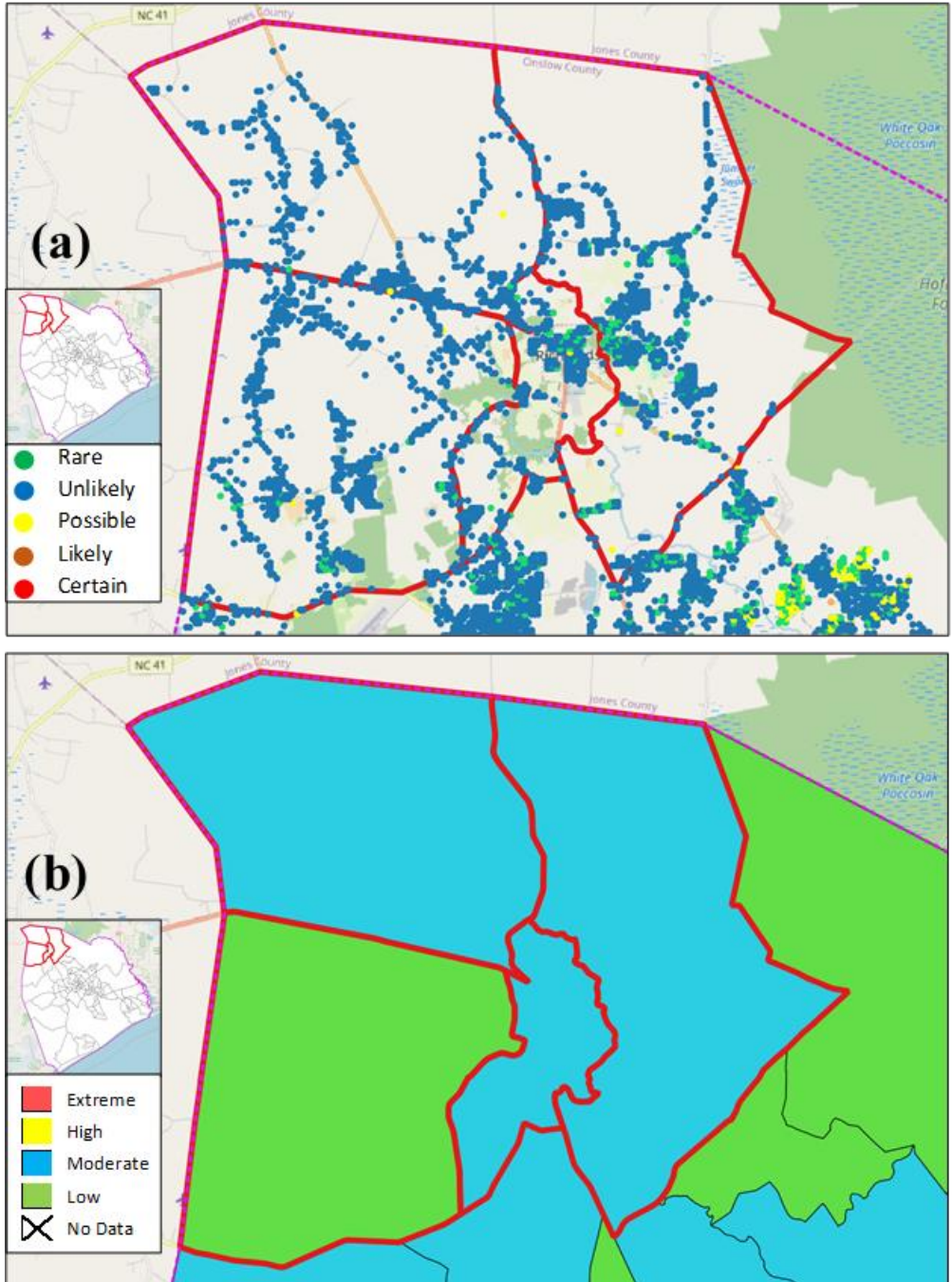

Figure 7. a) Spatial distributions of physical damage; b) overall risk, across the selected BGs

## 5 CLOSING REMARKS

The framework results in community-level wind risk maps at the specified resolution of the study area, which is the census block group level in this paper. The generated risk maps categorize the overall risk of wind-induced damages to residential buildings into one of four categories: low, moderate, high, and extreme. These maps are to aid community decision-makers in resilience and risk mitigation planning by providing a scale on how equity in such plans needs to be prioritized.

## 6 ACKNOWLEDGEMENTS

Research presented here was partially supported by the National Science Foundation under Grant No. CMMI 1847373. This material is also based upon work partially supported by the Center for Risk-Based Community Resilience Planning, funded through a cooperative agreement between the U.S. National Institute of Standards and Technology and Colorado State University (Grant Number 70NANB20H008). The content is solely the responsibility of the authors and does not necessarily represent the official views of the National Science Foundation or the National Institute of Standards and Technology.